\documentclass{PoS}

\usepackage{amsmath}
\usepackage{graphicx}
\graphicspath{{figs/}}

\newcommand{\comma}{\ ,}
\newcommand{\point}{\ .}
\newcommand{\Os}{\mathcal{O}_{\sigma}}
\newcommand{\Fpi}{F_{\pi}}
\newcommand{\Mpi}{M_{\pi}}
\newcommand{\Mro}{M_{\rho}}
\newcommand{\Msi}{M_{\sigma}}

\title{
\vspace*{-5cm}
{\normalsize {\rm \hfill{LLNL-PROC-666430} }} \\
\vspace*{5cm}
Conformality in twelve--flavour QCD}

\ShortTitle{Conformality in twelve--flavour QCD}

\author{
  Yasumichi Aoki$^a$\thanks{Presenter. E-mail: yaoki@kmi.nagoya-u.ac.jp},
  Tatsumi Aoyama$^a$,
  Ed Bennett$^b$,
  Masafumi Kurachi$^a$,
  Toshihide Maskawa$^a$,
  Kohtaroh Miura$^a$,
  Kei-ichi Nagai$^a$,
  Hiroshi Ohki$^a$, 
  Enrico Rinaldi$^c$\thanks{Presenter. E-mail: rinaldi2@llnl.gov},
  Akihiro Shibata$^d$,
  Koichi Yamawaki$^a$
  and 
  Takeshi Yamazaki$^e$ 
  
  \hspace*{55mm} (LatKMI Collaboration) 
  \\
  
  $^a$
  Kobayashi-Maskawa Institute for the Origin of Particles and the Universe (KMI), Nagoya University, Nagoya 464-8602, Japan \\
  $^b$
  Department of Physics, Swansea University, Singleton Park, Swansea SA2 8PP, UK \\
  $^c$ 
  Nuclear and Chemical Sciences Division, Lawrence Livermore National Laboratory, Livermore, CA 94550, USA \\
  $^d$
  Computing Research Center, High Energy Accelerator Research Organization (KEK), Tsukuba 305-0801, Japan \\
  $^e$
  Graduate School of Pure and Applied Sciences, University of Tsukuba, Tsukuba, Ibaraki 305-8571, Japan\\
}

\abstract{
The spectrum of twelve-flavor QCD has been studied in details by the LatKMI collaboration. In this proceeding we present our updated results for the spectrum obtained with the HISQ action at two lattice spacings, several volumes and fermion masses. In particular, we emphasize the existence of a flavor-singlet scalar state parametrically light with respect to the rest of the spectrum, first reported in our paper~\cite{Aoki:2013zsa}. This feature is expected to be present for theories in the conformal window, but the lattice calculation of such a state is difficult and requires noise-reduction techniques together with large statistics, in order to evaluate disconnected diagrams. Being able to provide a robust observed connection between a light flavor-singlet scalar and (near-)conformality is an important step towards observing a light composite Higgs boson in walking technicolor theories on the lattice~\cite{Aoki:2013ttt}. We also show updated results for the mass anomalous dimension $\gamma_m$ obtained from various spectral quantities, including the string tension, under the assumption that the theory is inside the conformal window.
}

\FullConference{The 32nd International Symposium on Lattice Field Theory,\\
		23-28 June, 2014\\
		Columbia University New York, NY}

\begin{document}

\section{Introduction}
\label{sec:introduction}

Four years ago the LatKMI collaboration started a program of lattice simulations and related theoretical work to systematically study SU(3) gauge theories with increasing number $N_f$ of massless flavors in the fundamental representation. Lattice simulations of such theories were aimed at the discovery of a near-conformal theory with a mass anomalous dimension $\gamma_m \sim 1$, a possible candidate for Walking Technicolor~\cite{Yamawaki:1985zg}.

For large number of flavors, the $\beta$ function of SU(3) gauge theories displays an infrared conformal fixed point in its perturbative expansion~\cite{BanksZaks}. This happens for a critical number of flavors $N_f > N_F^\star$, where $N_f^\star$ is close to $N_f=8$ for $2$-loop perturbation theory or $N_f=12$ for ladder Schwinger-Dyson analysis. The perturbative treatment is not completely justified at such value of $N_f^\star$ because the IRFP becomes strongly coupled. Non-perturbative lattice simulations are a suited tool for the study of this fixed point.

The SU(3) 12-flavor theory has been explored on the lattice by many different groups and using several different methods. LatKMI investigated signals of a IRFP in this theory by looking first at the spectrum of mesonic bound states and their decay constants, specifically the lightest pseudoscalar and vector state~\cite{Aoki:2012eq}. Subsequently, following the results of Ref.~\cite{DelDebbio:2009fd}, LatKMI investigated the gluonic spectrum, focusing on the scalar glueball state~\cite{Aoki:2013pca} and the string tension~\cite{Aoki:2013twa}. The encouraging results of Ref.~\cite{Aoki:2013pca} indicated the presence of a light flavor-singlet scalar state in the spectrum, lighter than the pseudoscalar one, and this state was carefully studied in another publication using fermionic operators as well as gluonic ones~\cite{Aoki:2013zsa}.

In this proceeding we report updated results to all of the aforementioned papers by LatKMI. The results presented in this publication are to be considered preliminary. The lattice simulations performed include larger volumes, smaller fermion masses and increased statistics with respect to published results. We use the Highly Improved Staggered Quark (HISQ) action and the tree--level Symanzik gauge action at two bare gauge coupling constants $\beta=6/g^2=3.7$ and $4.0$ and we simulate four physical volumes with $L=18, \, 24, \, 30, \, 36$ with aspect ratio $T/L = 4/3$. Different bare quark masses are used, going from $m_f = 0.2$ on the smallest $L=18$ volume, to $m_f=0.03$ on the largest $L=36$ ($m_f=0.04$ for $\beta=4.0$). As a reference, the simulated pseudoscalar (would--be pion) masses are $0.24 < \Mpi < 0.91$. Some of the features of these simulations, are the good flavor symmetry realization of the HISQ action and the large number of configurations, the latter needed for a precise determination of gluonic correlators and fermionic disconnected diagrams~\cite{Aoki:2013zsa}.

\section{Spectrum and finite-size scaling}
\label{sec:spectrum}

We first focus on the lightest flavor non-singlet mesonic resonances of the lattice theory. The pseudoscalar mass ($M_{\pi}$), the pseudoscalar decay constant ($F_{\pi}$) and the vector mass ($M_{\rho}$) are extracted from lattice Euclidean correlators of the corresponding staggered fermions operators. The standard techniques used are the same described in Ref.~\cite{Aoki:2012eq}. The goal is to look at this hadronic spectrum as a function of the bare fermion mass $m_f$ to identify signals of a conformal fixed point. If the theory is inside the conformal window the leading mass dependence of the spectrum is dictated by hyperscaling~\cite{hyperscaling} relations:
\begin{equation}
  \label{eq:hyperscaling}
  M_H \propto m_f^{\frac{1}{1+\gamma_m}} \comma \qquad F_H \propto m_f^{\frac{1}{1+\gamma_m}} \point
\end{equation}
Therefore dimensionless ratios of masses and decay constants should approach a constant value when the theory reaches the chiral limit where the IRFP is present. Moreover, we should remind that the relations in Eq.~(\ref{eq:hyperscaling}) are valid in the infinite volume limit and when the mass is the only source of soft breaking of scale invariance.

\begin{figure}[h]
  \centering
  \begin{tabular}{cc}
    \includegraphics[width=0.48\textwidth]{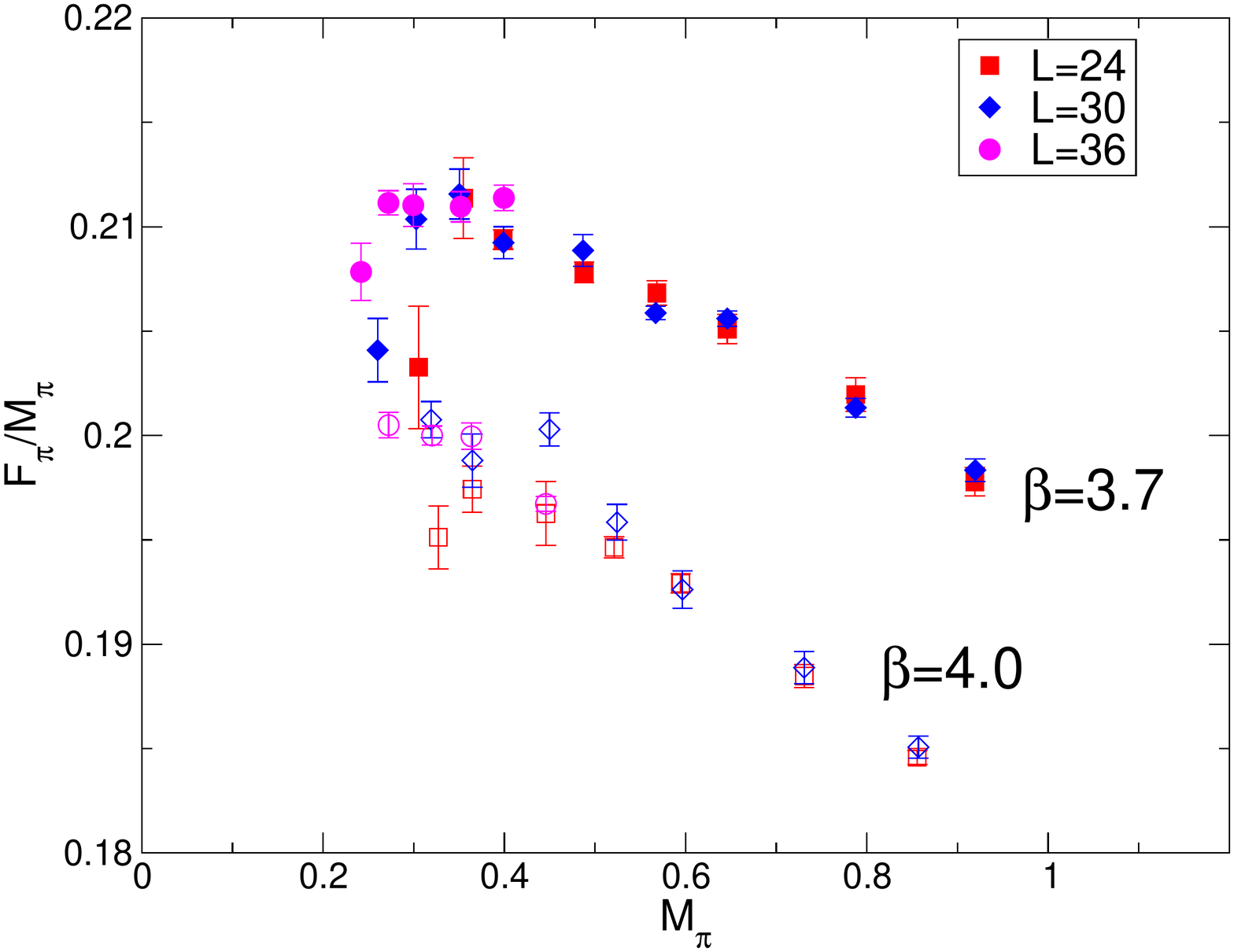} &
    \includegraphics[width=0.48\textwidth]{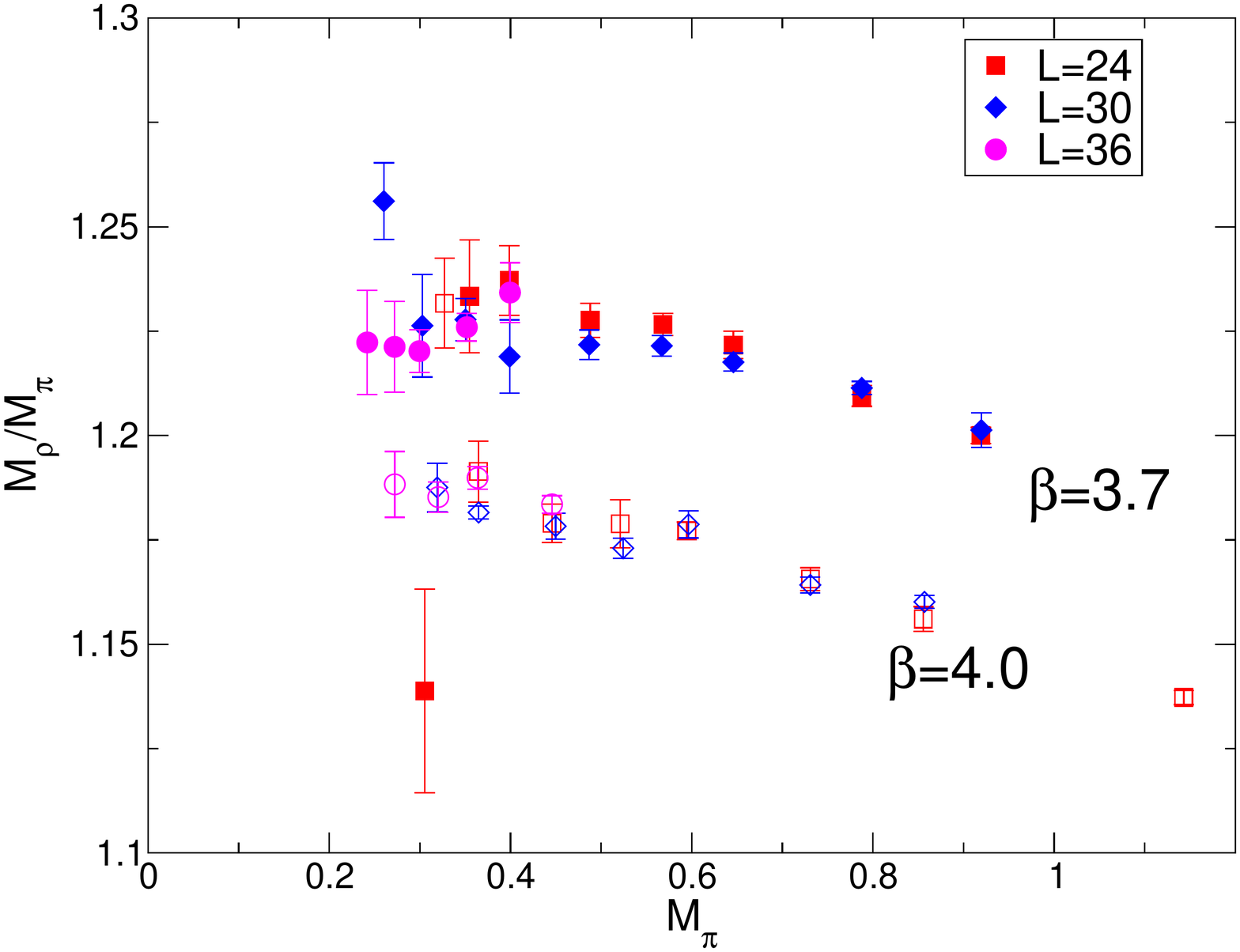} \\
  \end{tabular}
  \caption{(Left) The ratio of the pseudoscalar decay constant and pseudoscalar mass at two different values of the coupling constant. (Right) The ratio of the vector mass and the pseudoscalar mass at two different values of the coupling constant. In both cases the ratios approach an approximately constant value towards the chiral limit of the theory, signaling a common scaling function.}
  \label{fig:ratios}
\end{figure}

In Fig.~\ref{fig:ratios} we plot the dependence of two ratios, $\Fpi / \Mpi$ and $\Mro / \Mpi$, on $\Mpi$ as a proxy for the fermion mass $m_f$. For decreasing $\Mpi$ the theory is approaching the chiral limit and the dimensionless ratios first increase before reaching a plateaux value. These plateaux are only visible if larger volumes are considered at small pseudoscalar masses and this is a clear signal of how finite-volume effects can influence the system. The results presented in Fig.~\ref{fig:ratios} improve the similar ones in Fig.4 of Ref.~\cite{Aoki:2012eq} by adding data points on larger volumes and smaller pseudoscalar masses.

One can turn finite-volume effects, which distort hyperscaling relations, into a powerful tool to look for infrared conformality, in the same way one uses finite size scaling techniques to study critical points of finite systems in Statistical Mechanics. At finite fermion mass and finite lattice size, the presence of a conformal IRFP can be studied using universal scaling relations 
for hadronic masses and decay constants:
\begin{equation}
  \label{eq:fshs}
  \xi = LM_H = f_H(x) \comma  \qquad  \xi = LF_\pi = f_F(x) \comma
\end{equation}
where the subscript $H$ refers to a hadron ($\pi$ or $\rho$ here) and
the scaling variable is 
\begin{equation}
  \label{eq:ix}
  x=Lm_f^{\frac{1}{1+\gamma_m}} \point
\end{equation}

\begin{figure}[h]
  \centering
  \begin{tabular}{cc}
    \includegraphics[width=0.48\textwidth]{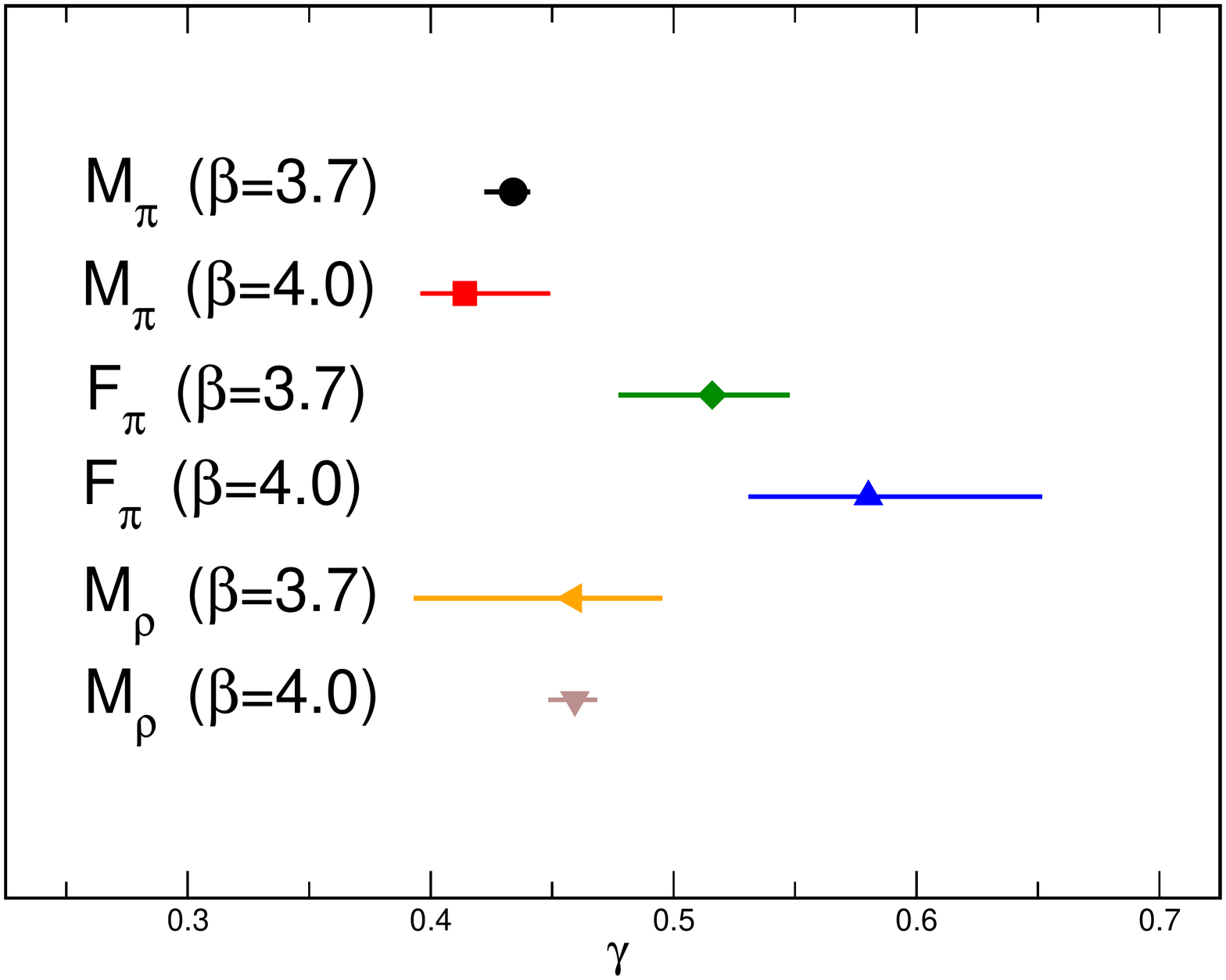} &
    \includegraphics[width=0.48\textwidth]{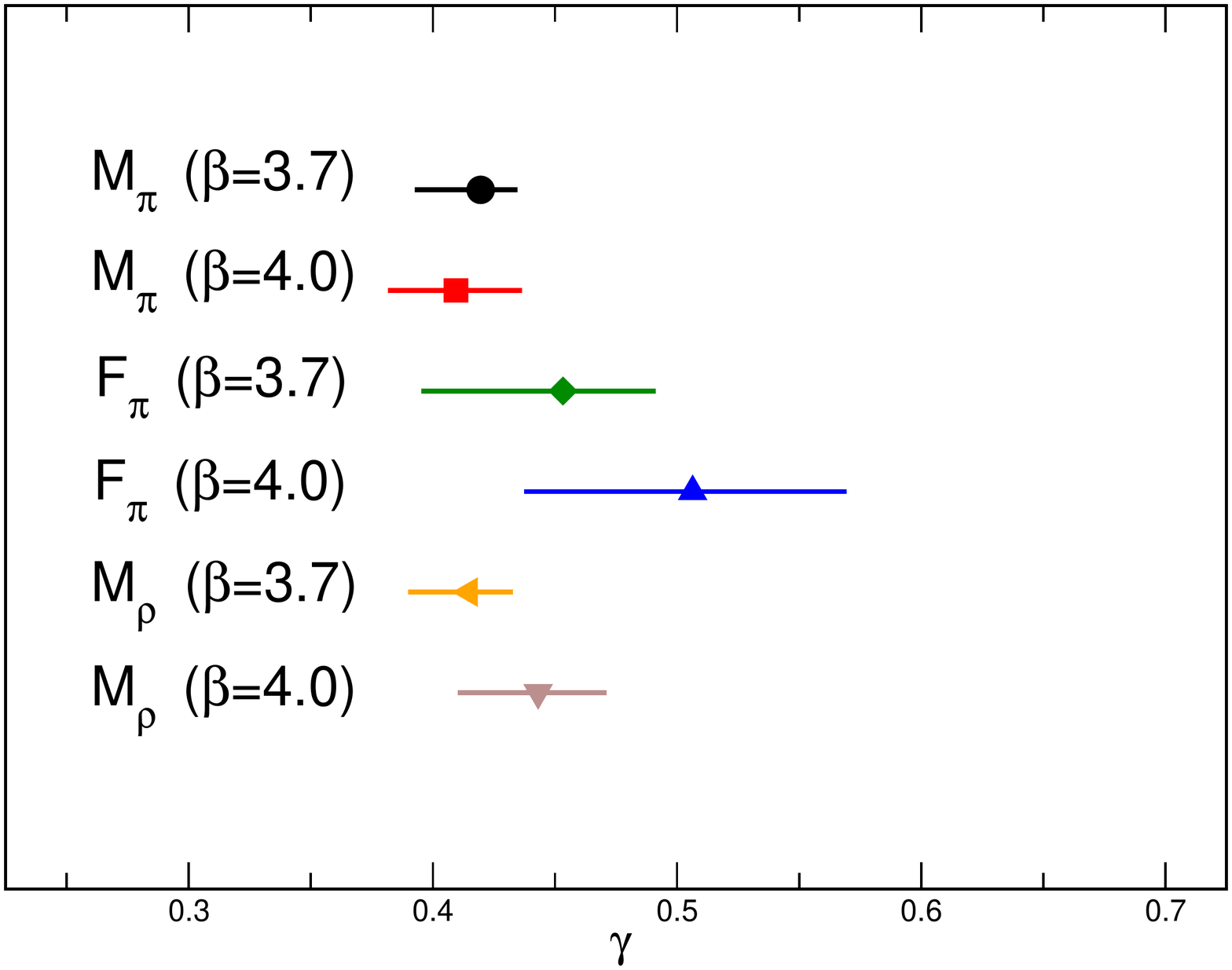} \\
  \end{tabular}
  \caption{(Left) Optimal value of $\gamma_m$ with statistical and systematic errors from Ref.~\cite{Aoki:2012eq}. (Right) Updated values of $\gamma_m$ for three different observables and two coupling constants using larger volumes at small fermion masses. The $\gamma_m$ obtained from the finite-size scaling analysis of $\Fpi$ is now compatible with the one coming from other channels.}
  \label{fig:gammas}
\end{figure}

In Ref.~\cite{Aoki:2012eq} the value of $\gamma_m$ is estimated by
quantifying the \emph{alignment} of the data points for $\xi(L,m_f)$ to
a universal function. The scaling functions 
$f_H(x)$ and $f_F(x)$
have to reach the asymptotic infinite-volume form $f_X(x) \sim x$ required by Eq.~(\ref{eq:hyperscaling}). Moreover, the value of $\gamma_m$ is characteristic of the critical point and must be independent of the observable used in the scaling analysis. The analysis presented in Ref.~\cite{Aoki:2012eq} showed that a common value of $\gamma_m$ could be extracted by looking at both $\Mpi$ and $\Mro$. However, the data for $\Fpi$ did hint at a $\gamma_m$ value that was not universal. This would weaken any possible claim about the existence of a conformal IRFP in the chiral limit of the theory. It is insightful hence to notice that including new data points on larger volumes drastically changes the situation. This can be seen by comparing the two panels of Fig.~\ref{fig:gammas}, corresponding to old published data~\cite{Aoki:2012eq} and newly available data presented for the first time in this proceeding. From the plot one can read of a value $0.4 \lesssim \gamma_m \lesssim 0.5$ for the anomalous mass dimension at the IRFP.

A different approach that includes corrections to the finite size hyperscaling relations can help in taking into account volume effects and non--universal corrections. For example, one could try and fit the data $\xi(L,m_f)$ using different functions:
\begin{align}
  \xi & =  c_0  + c_1 Lm_f^{1/(1+\gamma_m)} \label{eq:fitA}\\
  \xi & =  c_0  + c_1 Lm_f^{1/(1+\gamma_m)} + c_2Lm_f^\alpha \label{eq:fitB} \\
  \xi & =  (c_0  + c_1 Lm_f^{1/(1+\gamma_m)})(1+c_2m_f^\omega) \label{eq:fitC} \point
\end{align}
In Eq.~(\ref{eq:fitB}) the value of $\alpha$ is fixed to $\alpha=(3-2\gamma_m)/(1+\gamma_m)$, inspired by solution of the Schwinger-Dyson equation~\cite{sdLatKMI}, or $\alpha=2$, regarded as a lattice discretization artefact. In Eq.~(\ref{eq:fitC}), the additional term $\propto m_f^\omega$ is inspired by the work in Ref.~\cite{Cheng:2013xha}, where corrections due to the (near-)marginal gauge coupling operator have been shown to be relevant in certain parameter regions of the simulations. By using fit forms above on large volume data ($L\Mpi>8.5$ and $L\Fpi>2$) the fitted value $\gamma_m$ turns out to be consistent across different observables and coupling constants, as well as consistent with the previous analysis shown in Fig.~\ref{fig:gammas} (Right). Including corrections due to near-marginal operators near the fixed point ($\omega \neq 0$) gives compatible results and it also improves the quality of the fits in the heavy mass region.

\section{Flavor-singlet scalar}
\label{sec:iso-singlet-scalar}

When a conformal theory is perturbed by a mass deformation as in our lattice simulations at fixed $m_f$, a flavor-singlet scalar excitation in the spectrum may be parametrically light with respect to the other resonances. This non-perturbative physical phenomenon was observed first on the lattice using gluonic interpolating operators in the SU(2) theory with two adjoint flavors~\cite{DelDebbio:2009fd}, and then observed also in SU(3) with twelve fundamental flavors~\cite{Aoki:2013pca,Aoki:2013zsa} using, for the first time, fermionic correlators. We employ large computational resources to measure disconnected fermionic contributions that are necessary for the correct analysis of the flavor-singlet scalar correlator.

In order to study the flavor-singlet scalar correlator and its ground state, which we call $\sigma$ in analogy to QCD, we employ the local fermionic bilinear operator 
\begin{equation}
  \label{eq:fermionic-scalar-op}
  \Os(t) \; = \; \sum_{i=1}^{3}\sum_{\vec{x}} \overline{\chi}_i(\vec{x},t) \chi_i(\vec{x},t) \ ,
\end{equation}
where the index $i$ runs through different staggered fermion species. Then we calculate its correlator, which is constructed by both the connected $C(t)$ and vacuum--subtracted disconnected $D(t)$ correlators, $\langle \Os(t) \Os^\dag(0) \rangle = 3 D(t) - C(t)$, where the factor in front of $D(t)$ comes from the number of species. Following the analysis performed in Ref.~\cite{Aoki:2013zsa} we extract an effective mass for the state in the scalar channel and we study its dependence on the bare fermion mass in order to check if hyperscaling takes place. Due to the higher computational cost we focus on a fixed lattice spacing $\beta=4.0$.

\begin{figure}[h]
  \centering
  \begin{tabular}{cc}
    \includegraphics[width=0.48\textwidth]{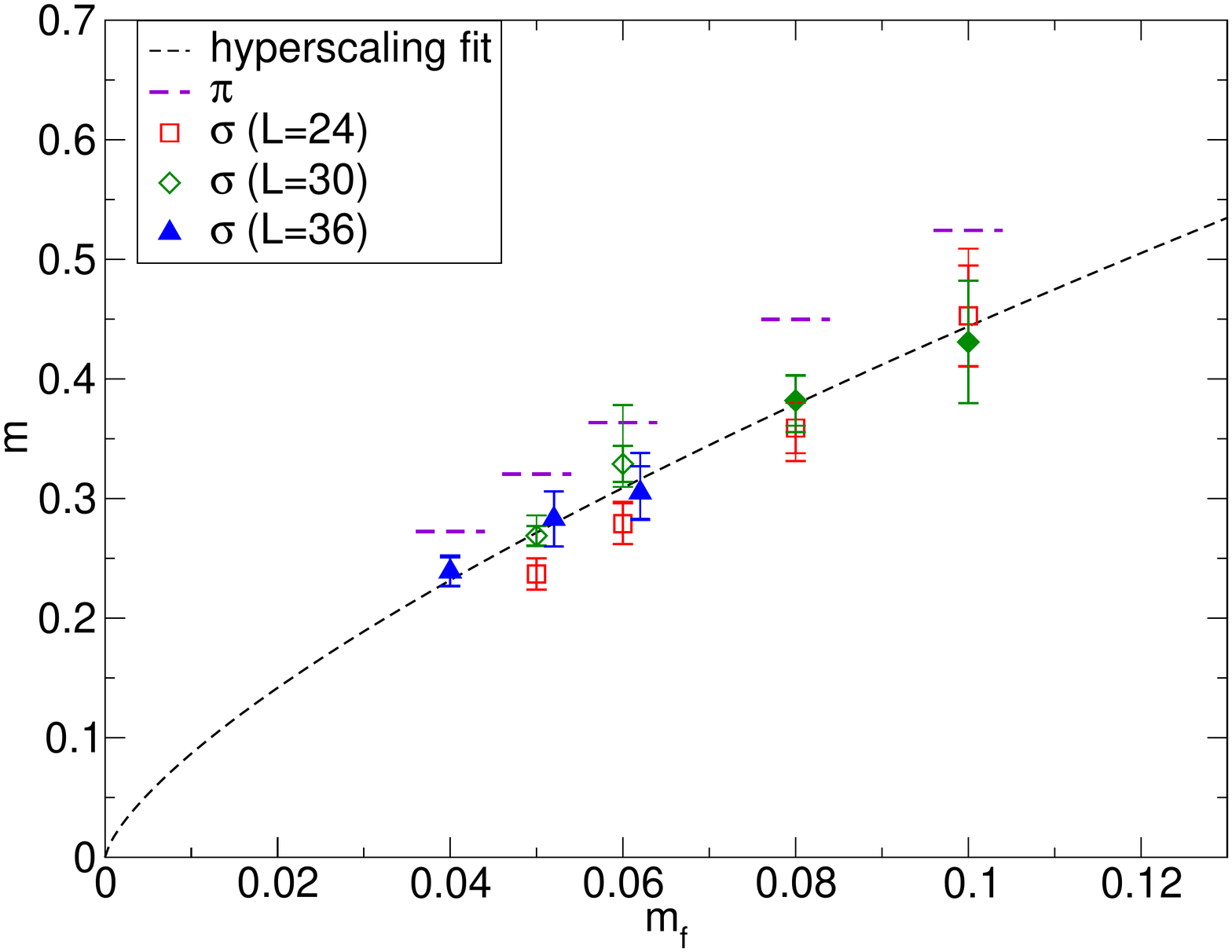} &
    \includegraphics[width=0.48\textwidth]{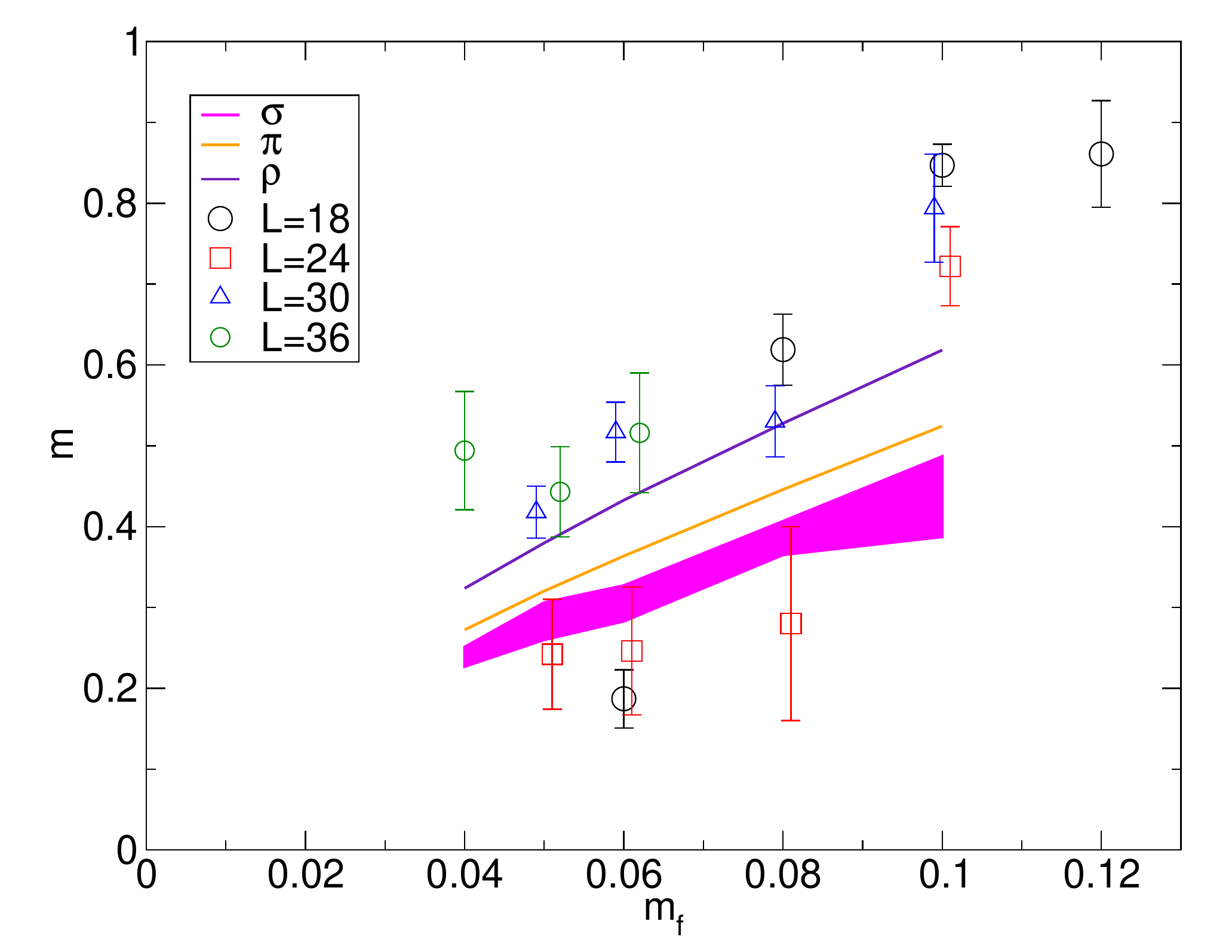} \\
   \end{tabular}
  \caption{(Left) Spectrum of the flavor-singlet scalar obtained on different volumes from fermionic interpolating operators. The pseudoscalar spectrum is shown for reference. Only the filled symbols are included in the hyperscaling fit. (Right) The spectrum in the scalar channel obtained from gluonic interpolating operators. $\Mpi$, $\Mro$ and $\Msi$ are all shown for reference.}
  \label{fig:scalars}
\end{figure}

The left panel of Fig.~\ref{fig:scalars} shows that the lightest scalar state is lighter than the pseudoscalar one in the whole mass region explored. The fermion mass dependence is well described by Eq.~(\ref{eq:hyperscaling}) when $\gamma_m$ is fixed to the value extracted from the study of $\Mpi$. The right panel of Fig.~\ref{fig:scalars} instead shows the comparison with the scalar spectrum obtained from gluonic interpolation operators (following the methodology of Ref.~\cite{Lucini:2010nv}). A scalar glueball seems to exist in the spectrum, but it is at the same scale of the vector meson.

\section{String tension}
\label{sec:string-tension}

\begin{figure}[h]
  \centering
  \includegraphics[width=0.65\textwidth]{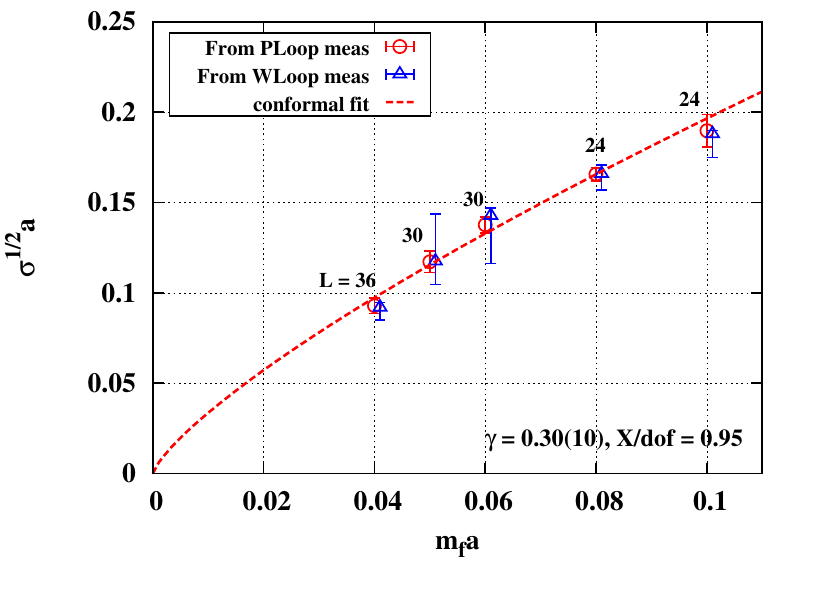}
  \caption{String tension values in units of the lattice spacing for $\beta=4.0$. Estimates obtained from Polyakov loop correlators and rectangular Wilson loop correlators are shown to be compatible for a range of volumes and masses. The Polyakov loops estimate is fitted with a hyperscaling form and gives $\gamma_m=0.3(1)$, broadly compatible with the rest of the hadronic spectrum.}
  \label{fig:string}
\end{figure}

We also study purely gluonic quantities to extend the hyperscaling analysis to more observables, which are affected by different systematic effects compared to the fermionic ones~\cite{Aoki:2013twa}. Among the gluonic observables explored in the SU(3) $N_f=12$ theory is the string tension $\sqrt{\sigma}$. Using two different type of correlators, we obtain compatible results for the string tension at $\beta=4.0$ and for several fermion masses and volumes. The first method to obtain $\sqrt{\sigma}$ is from smeared Polyakov loop correlators coupling to torelon excitations whose mass grows with the length of the loop
\begin{equation}
  \label{eq:sigma}
  m_{\rm tor}(L)  =  \sigma L - \frac{\pi}{3L} - \frac{\pi^2}{18L^3}\frac{1}{\sigma} \point
\end{equation}
The second method uses APE smeared Wilson loop operators to construct Creutz ratios which describe the static quark-antiquark potential
\begin{equation}
  \label{eq:wilson}
 V(r) =  v_0 - \frac{\alpha}{r} + \sigma^2 r \point
\end{equation}
In Fig.~\ref{fig:string} we show how well the two methods above can extract $\sqrt{\sigma}$: a result obtained from methodologies that have different systematic errors is very robust. The same plot also displays a hyperscaling fits of the data obtained from the Polyakov loop correlators. The fit has a very good $\chi^2/\textrm{dof}$ and the fitted mass anomalous dimension $\gamma_m=0.3(1)$ is compatible with the rest of the spectrum.

\section{Conclusions}
\label{sec:conclusions}

Finite-size hyperscaling formulae describe the lattice data for the pseudoscalar mass and decay constant and for the vector mass, resulting in a universal value for the anomalous mass dimension. Both the flavor-singlet scalar mass and the string tension are compatible with these predictions. 
These are indications that the twelve-flavor SU(3) gauge theory simulated on the lattice corresponds to a conformal theory in the continuum and infinite volume limit.
We also note that chiral perturbation theory does not apply to this context for the reason that the expansion parameter is much larger than unity and that the would--be pions are no longer the lightest degrees of freedom of the theory.

Other theories that are thought to be close to or inside the conformal window seem to have a spectrum that contains a relatively light flavor-singlet scalar particle. If this feature continues to hold as the lattice studies become more precise and explore larger volumes at smaller fermion masses, it will be important to develop a fully fledged effective field theory that can take into account a scalar degree of freedom interacting together with the pseudoscalar ones. This is relevant for Walking Technicolor models, Composite Higgs models and any theory with light scalars.

\vspace{0.3cm}
{\it Acknowledgments --} Numerical calculations have been carried out on
 the high-performance computing systems at KMI ($\varphi$), at 
 the Information Technology Center in Nagoya University (CX400), and 
 at the Research Institute for Information
 Technology in Kyushu University (CX400 and HA8000).
 This work is supported by the JSPS Grant-in-Aid for Scientific Research
 (S) No.22224003, (C) No.23540300 (K.Y.), for Young Scientists (B)
 No.25800139 (H.O.) and No.25800138 (T.Y.), and also by the MEXT
 Grants-in-Aid 
 for Scientific Research on Innovative Areas No.23105708
 (T.Y.). E.R. acknowledges the support of the U.S. Department of Energy
 under Contract \ DE-AC52-07NA27344 (LLNL).


\end{document}